\documentclass[aps,prl,twocolumn,superscriptaddress]{revtex4-1}
\usepackage{amsfonts, amssymb, amsmath,graphicx}
\usepackage{subfigure}
\usepackage{multirow}
\usepackage{dcolumn}
\usepackage{bm}
\usepackage{color}

\begin{document}

\title{Probing localization and quantum geometry by spectroscopy}

\author{Tomoki Ozawa}
\affiliation{Interdisciplinary Theoretical and Mathematical Sciences Program (iTHEMS), RIKEN, Wako, Saitama 351-0198, Japan}
\author{Nathan Goldman}
\affiliation{Center for Nonlinear Phenomena and Complex Systems, Universit\'e Libre de Bruxelles, CP 231, Campus Plaine, B-1050 Brussels, Belgium}%

\date{\today}

\newcommand{\tom}[1]{{\color{red} #1}}
\definecolor{Nathanblue}{rgb}{0.,0.24,0.51}
\newcommand{\blue}{\color{Nathanblue}}

\begin{abstract}
The spatial localization of quantum states plays a central role in condensed-matter phenomena, ranging from many-body localization to topological matter. Building on the dissipation-fluctuation theorem, we propose that the localization properties of a quantum-engineered system can be probed by spectroscopy, namely, by measuring its excitation rate upon a periodic drive. We apply this method to various examples that are of direct experimental relevance in ultracold atomic gases, including Anderson localization, topological edge modes, and interacting particles in a harmonic trap. Moreover, inspired by a relation between quantum fluctuations and the quantum metric, we describe how our scheme can be generalized in view of extracting the full quantum-geometric tensor of many-body systems. Our approach opens an avenue for probing localization, as well as quantum fluctuations, geometry and entanglement, in synthetic quantum matter. 
\end{abstract}

\maketitle

Localization plays a central role in various branches of quantum physics. In the context of the solid state, the localization properties of electronic wave functions reflect the conductivity of materials and signal the existence of insulating regimes~\cite{Anderson:1958,Lee:1985, Kramer:1993, Beenakker:1997,Kudinov:1991,Resta:1999}. This important relation between transport and the spatial localization of quantum states was already emphasized in the seminal work by Anderson on disordered lattice systems~\cite{Anderson:1958}. More recently, the discovery of topological states of matter revealed an interesting interplay between topology and localization:~The bulk-boundary correspondence guarantees the existence of robust boundary modes, which are localized at the edge of the material~\cite{Hasan:RMP, Qi:RMP}. Another important development concerns the phenomenon of many-body localization (MBL), which is characterized by the absence of thermalization in many-body systems featuring disorder and interparticle interactions~\cite{Basko:2006,Nandkishore:2015,Abanin:2018}.

Quantum engineered systems, such as ultracold atomic gases or trapped ions, have recently emerged as novel platforms by which localization can be finely studied in a highly controlled environment. Anderson localization was observed in Bose-Einstein condensates, in various spatial dimensions, through the design of disordered potentials for neutral atoms~\cite{Billy:2008, Roati:2008, Kondov:2011}. Besides the control over the disorder strength, these quantum-engineered systems also allow for the possibility of tuning the interparticle interactions. Combining these two appealing features led to the first experimental observations of MBL in ultracold atomic gases~\cite{Schreiber:2015, Choi:2016}, which were soon followed by realizations in trapped ions~\cite{Smith:2016} and in photonics~\cite{Roushan:2017}. While the localization length was directly measured in the context of Anderson localization~\cite{Billy:2008,Kondov:2011}, by imaging the spatial profile of the atomic cloud \textit{in situ}, such direct signatures of localization remain challenging in the MBL regime (see Ref.~\cite{Greiner_MBL} for correlation-length measurements in the MBL phase based on single-site resolution imaging).

In this Rapid Communication, we introduce a novel method by which localization can be quantitatively studied in quantum-engineered systems, without relying on any {\it in situ} imaging. Our approach builds on a universal relation between the localization of a quantum state and its spectroscopic response to an external periodic drive and therefore can be applied to many-body (interacting) systems. In fact, this relation between localization properties and dissipative responses can be traced back to the fluctuation-dissipation theorem~\cite{Callen:1951, Kubo:1957, LandauLifshitz:Book}, as was previously noticed in the solid state~\cite{Kudinov:1991,Souza:2000}. In this Rapid Communication we propose that measuring the excitation rate of a quantum-engineered system upon a periodic drive offers a practical scheme by which its localization properties can be precisely evaluated. In particular, this method can be readily applied to general many-body systems, in the presence of interactions and/or disorder, as we illustrate below through relevant examples. Our proposal builds on the general observation that excitation-rate measurements can be used to extract useful information on the underlying quantum states, as was recently illustrated in the context of topological Bloch bands~\cite{Tran:2017,Tran:2018,OzawaGoldman, Asteria:2018}.

Beyond the localization of particles in extended lattice geometries, our method can also be applied to study localization in confined systems. This asset is relevant to ultracold atomic gases in optical lattices, where the spread of the two-body wavefunction within each lattice site affects the effective interaction energy~\cite{Campbell:2006, Will:2010,Will:2011,Zurn:2012}. While detecting localization within a single site of an optical lattice requires sophisticated nanoscale microscopy~\cite{sub_resolution,sub_resolution_2}, this property could be equally studied using the more practical spectroscopic probe introduced in this work.

Dissipation and fluctuations are united through the imaginary part of the generalized susceptibility~\cite{Callen:1951, Kubo:1957, LandauLifshitz:Book}. 
Interestingly, it was noted that this response function is also deeply related to the geometry of quantum states, through the notion of the quantum (Fubini-Study) metric~\cite{Provost:1980, Souza:2000,Kolodrubetz:PhysRep} (see Ref.~\cite{Hauke_Heyl}  for a similar observation involving the quantum Fisher information, which captures the entanglement properties of many-body quantum states). In many-body systems, the quantum metric is defined over the parameter space spanned by twist angles associated with boundary conditions~\cite{Souza:2000}. As a byproduct of our proposal, we establish a protocol by which the full quantum geometry of many-body quantum systems (including the quantum metric and the many-body Berry curvature~\cite{Niu:1985,Watanabe:2018,Kudo:2019}) can be extracted from spectroscopic responses.

\textit{Excitation rates and localization.}
We first discuss how the excitation rate of a quantum system upon a periodic drive relates to the variance of the position operator. Let us assume that the system is initially in an eigenstate $|\alpha\rangle$ of a many-body Hamiltonian $\hat{H}$, which we consider to be non-degenerate and well separated from all other states in energy~\cite{remark0}. We then act on the system with a time-periodic perturbation aligned along the $x$ direction so that the total Hamiltonian reads
\begin{align}
	\hat{H} (t)
	=
	\hat{H}
	+
	2E \hat{x} \cos (\omega t),\label{driving_ham}
\end{align}
where $\hat{x}\!\equiv\!\sum_a \hat{x}_a$ is the many-body position operator along the $x$ direction and $\hat{x}_a$ is the position operator for the $a$th particle.
At the lowest order in time-dependent perturbation theory, the excitation fraction is given by Fermi's golden rule~\cite{Sakurai}
\begin{align}
	n_{\hat{x}}^\mathrm{ex}(\omega, t)
	=
	\frac{2\pi E^2}{\hbar} t \sum_{\beta \neq \alpha}
	\left| \left\langle \beta \left|
	\hat{x}
	\right|\alpha \right\rangle \right|^2
	\delta^{(t)} (\epsilon_\beta - \epsilon_\alpha - \hbar \omega),\notag
\end{align}
where $\delta^{(t)} (\epsilon) \equiv (2\hbar / \pi t)\sin^2 (\epsilon t/2\hbar)/\epsilon^2$ can be approximated by a Dirac distribution for sufficiently large $t$.
It is convenient to introduce the excitation rate $\Gamma_{\hat{x}} (\omega)\!\equiv\!n_{\hat{x}}^\mathrm{ex}(\omega, t)/t$, which is the rate at which the system excites to other states $\beta\!\ne\!\alpha$. 
Inspired by Refs.~\cite{Tran:2017,Tran:2018,OzawaGoldman}, we consider performing a set of experiments for various values of the shaking frequency $\omega$, and integrating the resulting rates over $\omega$, 
\begin{align}
	\Gamma_{\hat{x}}^{\mathrm{int}}
	&\equiv
	\int_0^\infty \Gamma_{\hat{x}} (\omega) d\omega 
	=
	\frac{2\pi E^2}{\hbar^2}\sum_{\beta \neq \alpha}
	\left| \left\langle \beta \left|
	\hat{x}
	\right|\alpha \right\rangle \right|^2.\label{eq_int}
\end{align}
Using the completeness relation $\sum_\beta |\beta\rangle \langle \beta |\!=\!1$, we rewrite the sum above as
\begin{align}
	\sum_{\beta \neq \alpha}
	\left| \left\langle \beta \left|
	\hat{x}
	\right|\alpha \right\rangle \right|^2
	&=
	\sum_{\beta \neq \alpha}
	\langle \alpha | \hat{x}| \beta\rangle \langle \beta | \hat{x} |\alpha\rangle
	\notag \\
	&=
	\left(
	\langle \alpha | \hat{x}^2 |\alpha\rangle - \langle \alpha | \hat{x}| \alpha\rangle^2
	\right)
	\equiv \mathrm{Var}(\hat{x}),
\end{align}
which is nothing but the variance $\mathrm{Var}(\hat{x})$ of the operator $\hat{x}$.
When combined with Eq.~\eqref{eq_int}, this yields a relation between the integrated rate and the spatial variance 
\begin{align}
	\Gamma_{\hat{x}}^{\mathrm{int}}
	=
	\frac{2\pi E^2}{\hbar^2}
	\mathrm{Var}(\hat{x}) .
	\label{integrated_rate}
\end{align}
This relation establishes a protocol by which the variance of the position operator can be measured in experiments, without detecting the position of the particles.
 
In fact, the relation between the integrated rate and the variance of an operator can be made more general. For any operator $\hat{A}$, the excitation rate upon the drive $\hat V (t)\!=\!2E \hat{A} \cos (\omega t)$ is related to the variance of this operator  through $\Gamma^{\mathrm{int}}_{\hat{A}}\!=\!\frac{2\pi E^2}{\hbar^2} \mathrm{Var}(\hat{A})$. Hence, time modulation can be used as a universal probe for the variance of any operator.
Moreover, noting that the variance of an operator is directly related to a quantum Fisher information~\cite{Hauke_Heyl}, we point out that this probe also captures the multipartite entanglement properties of many-body quantum states.

Excitation rates can be measured in various ways in quantum-engineered systems (see Refs.~\cite{Aidelsburger2015,Jotzu2015,Reitter,Boulier,Asteria:2018} for distinct experimental methods applied to periodically driven atomic gases). In optical lattices, heating rates can be finely measured by monitoring the dynamical repopulation of Bloch bands through band mapping~\cite{Aidelsburger2015,Asteria:2018}. In cases where intraband transitions are relevant and difficult to resolve~\cite{Repellin:2018}, one can optimize such measurements by applying a drive that simultaneously changes the internal state of the atoms~\cite{GoldmanBeugnonGerbier}; the excited fraction can then be determined by counting the number of atoms in  different internal states.

We point out that the result in Eq.~\eqref{integrated_rate} derives from the fluctuation-dissipation theorem. To see this, recall that the fluctuation associated with an operator $\hat{A}$, in a given state $|\alpha\rangle$, is related to the generalized susceptibility $\chi_{\hat{A}\hat{A}}(\omega)$ via the fluctuation-dissipation theorem~\cite{LandauLifshitz:Book}
\begin{align}
	\langle \alpha | \hat{A}^2 |\alpha\rangle
	=
	\frac{\hbar}{\pi}\int_0^{\infty} d\omega\, \mathrm{Im}\left[ \chi_{\hat{A}\hat{A}}(\omega)\right],
\end{align}
where we assumed $\langle \alpha |\hat{A}|\alpha\rangle\!=\!0$. In addition, the power absorbed $\mathcal{P}(\omega)$ upon a periodic drive $2E \hat{A} \cos (\omega t)$ is related to the imaginary part of the generalized susceptibility~\cite{LandauLifshitz:Book}: $ \mathcal{P}(\omega) = 2 \omega E^2 \mathrm{Im}\left[ \chi_{\hat{A}\hat{A}}(\omega)\right]$.
Noting that the excitation rate is defined as $\Gamma(\omega)\!=\!\mathcal{P}(\omega)/\hbar \omega$, we recover the relation $\Gamma^{\mathrm{int}}_{\hat{A}}\!=\!\frac{2\pi E^2}{\hbar^2} \mathrm{Var}(\hat{A})$.

\begin{figure}[h!]
\begin{center}
\includegraphics[width=0.7\columnwidth]{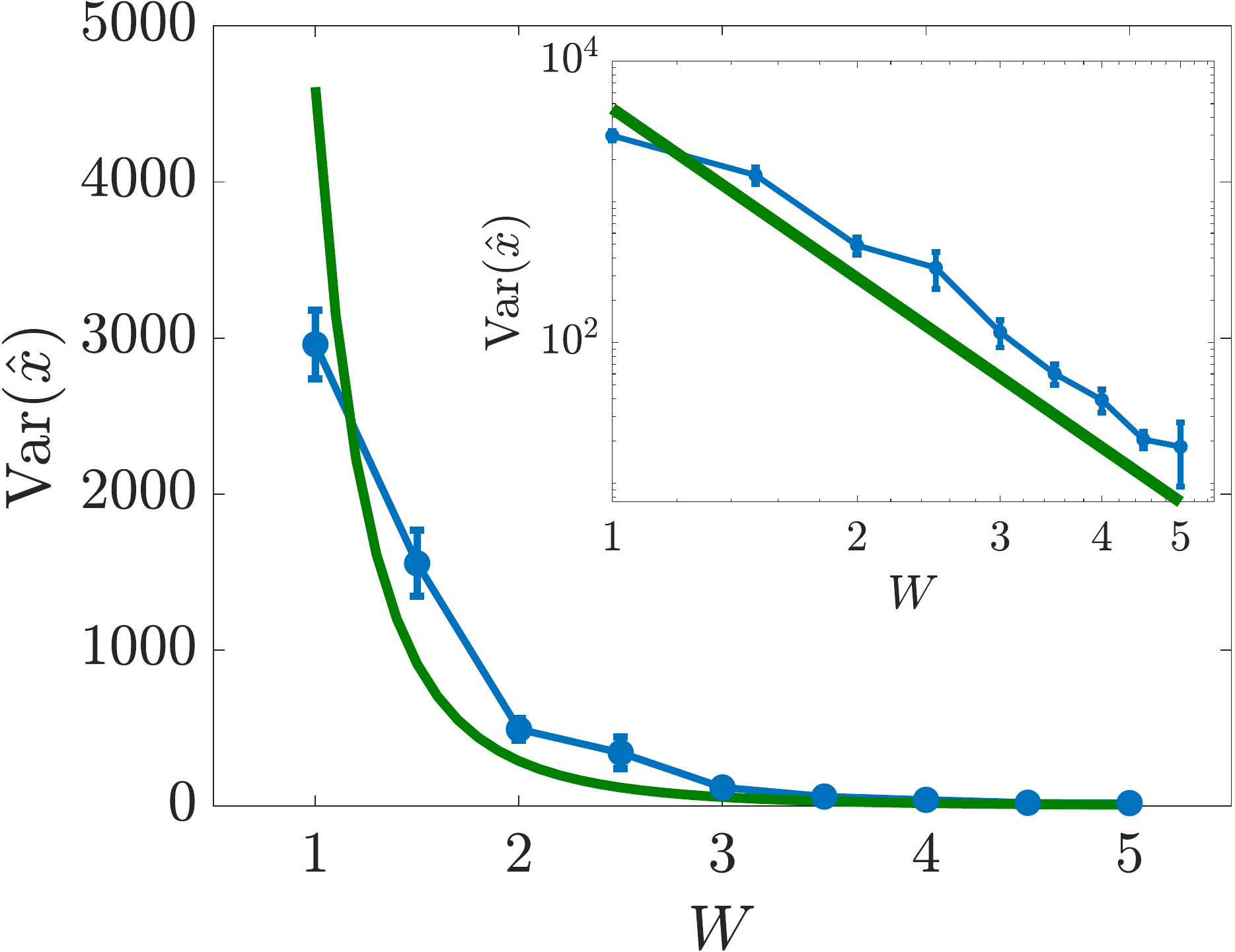}
\caption{Spatial variance $\mathrm{Var}(\hat{x})$ of zero-energy eigenstates as a function of the disorder strength $W$, as extracted from excitation-rate measurements [Eq.~\eqref{integrated_rate}]. Blue dots are numerical results obtained from 300 sites and an averaging over 20 disorder realizations (for each $W$). We used a modulation strength $E\!=\!0.001 J$ and an observation time of $t\!=\!5\hbar /J$; the excitation rates were integrated over $\omega$ up to the value $5J/\hbar$, using  discrete steps of $0.01J/\hbar$. The scaling-theory prediction $\mathrm{Var}(\hat{x})\!\approx \!4608J^4/W^4$ is also displayed (green curve). 
}
\label{disorder}
\end{center}
\end{figure}

\textit{Anderson model.}
We first apply our method to the Anderson model~\cite{Lee:1985,Kramer:1993}, which describes a quantum particle moving in a one-dimensional disordered lattice; the hopping matrix element is denoted $J$ and the random (disordered) potential has values uniformly distributed within the interval $[-W/2,W/2] J$. 
The presence of disorder generates localized eigenstates:~The envelope of the wave functions takes the form $\sim e^{-|x|/\xi}$ around their average position, where $\xi$ is the localization length. The spatial variance calculated from this naive exponentially decaying wave function reads $\mathrm{Var}(\hat{x})\!=\!\xi^2/2$, which indeed provides a qualitatively accurate estimate (deviations due to finer structures in the wavefunctions are illustrated below). According to the scaling theory of the Anderson model~\cite{Kramer:1993}, the localization length at zero energy scales as $\xi\!\approx\!96J^2/W^2$, in units of the lattice spacing; the corresponding variance reads $\mathrm{Var}(\hat{x})\!\approx\!4608J^4/W^4$ (see Fig.~\ref{disorder}).

We now show that these localization properties can be detected through excitation-rate measurements [Eq.~\eqref{integrated_rate}]. For a given disorder realization, we start from an eigenstate of the model whose energy is close to zero and then numerically calculate the integrated excitation rate upon applying a periodic modulation [Eq.~\eqref{driving_ham}]. The results are shown in Fig.~\ref{disorder}, for various values of the disorder strength $W$, together with the scaling-theory prediction. Fitting the estimated variance with a power law, we obtain $\mathrm{Var}(\hat{x})\!\approx\!4700 /W^{3.4}$, which is indeed very close to the prediction $\sim 1/W^4$; the main discrepancy is attributed to the finer structure of the wave function inside the envelop function $e^{-|x|/\xi}$.
We note that the eigenstate considered in these calculations is not isolated in energy (there is no spectral gap in the thermodynamic limit); however, states with similar energies are well separated spatially, and hence they do not contribute to the excitation rate~\cite{remark0}.

\begin{figure}[htbp]
\begin{center}
\includegraphics[width=1.0\columnwidth]{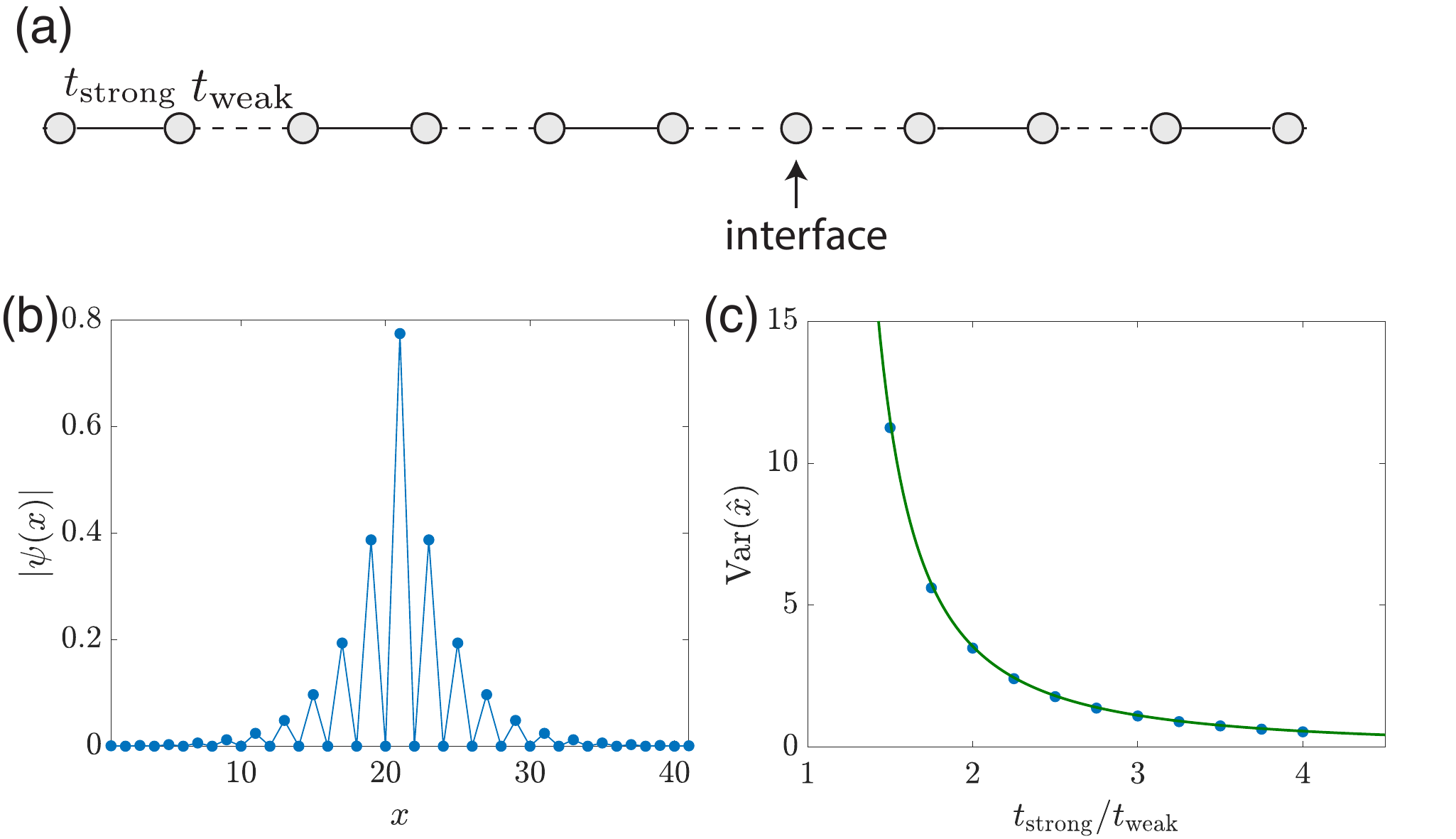}
\caption{(a) Schematics of the Su-Schrieffer-Heeger model, a one-dimensional chain with alternating hopping strengths. We consider a situation where a topological interface is created in the middle of the chain, by introducing a defect (where hoppings are weak $t_\mathrm{weak}$ on either side of a site). In this way, we obtain a zero-energy state that is localized at the interface. (b) Typical zero-energy mode localized at the interface. We used $t_\mathrm{strong}/t_\mathrm{weak} = 2$. The interface is located at site 21, where the wave function is maximal. (c) Variance $\mathrm{Var}(\hat{x})$ estimated from excitation-rate measurements upon linear driving, and the analytical value (\ref{analytical}) as a function of $t_\mathrm{strong}/t_\mathrm{weak}$. We used $E = 0.01 t_\mathrm{weak}/a$ and an observation time $t\!=\!5\hbar/t_\mathrm{weak}$; the excitation rates were integrated over $\omega$ up to the value $5 t_\mathrm{weak}/\hbar$, using  discrete steps of $0.05t_\mathrm{weak}/\hbar$.}
\label{sshfigure}
\end{center}
\end{figure}

\textit{Topological edges modes.} As a second example, we consider the celebrated Su-Schrieffer-Heeger (SSH) model, a model exhibiting symmetry-protected topological edge modes~\cite{Xiao:RMP,Cooper:RMP}. This is a one-dimensional tight-binding model with alternating hopping strengths $t_\mathrm{weak}$ and $t_\mathrm{strong}$; here we assume $t_\mathrm{weak} < t_\mathrm{strong}$. Depending on the termination of the chain, the boundaries can host localized (zero-energy) edge modes; these edge states are protected by topology, due to the chiral symmetry of the model~\cite{Hatsugai:2006}. Importantly, topological modes can also appear in the bulk of the chain~\cite{Leder:2016} whenever the latter presents a defect (e.g., if the strengths of the hopping amplitudes $t_\mathrm{weak}$ and $t_\mathrm{strong}$ are locally interchanged); such a defect constitutes an interface between two regions associated with different topological invariants (winding numbers), which explains the presence of a zero-energy mode that is exactly localized at the interface [see a sketch of the setting and the corresponding localized wave function depicted in Figs.~\ref{sshfigure}(a) and~\ref{sshfigure}(b)]. We note that this wavefunction only takes nonzero values on every other site. Analytically, the overall decay of the wave function obeys $|\psi(x)|\!=\!(t_\mathrm{weak}/t_\mathrm{strong})^{x/2a}$, where $x$ is the distance from the interface and $a$ is the lattice spacing. 
The spatial variance calculated from this analytical wavefunction reads
\begin{align}
\mathrm{Var}(\hat{x}) = 8 (t_\mathrm{weak}/t_\mathrm{strong})^2/[1-(t_\mathrm{weak}/t_\mathrm{strong})^2]^2. \label{analytical}
\end{align}
We also assume that both ends of the chain are terminated with a strong link ($t_\mathrm{strong}$), so there is no additional edge state at the boundaries of the chain; in this setting, we have a single zero-energy mode, which is localized at the interface. We performed simulations of the driven SSH model, in view of numerically estimating $\mathrm{Var}(\hat{x})$ from the integrated excitation rates. The results are plotted in Fig.~\ref{sshfigure}(c), where excellent agreement is shown between the excitation-rate measurement (dots) and the analytical result (solid line). 

\begin{figure}[htb]
\begin{center}
\includegraphics[width=1.0\columnwidth]{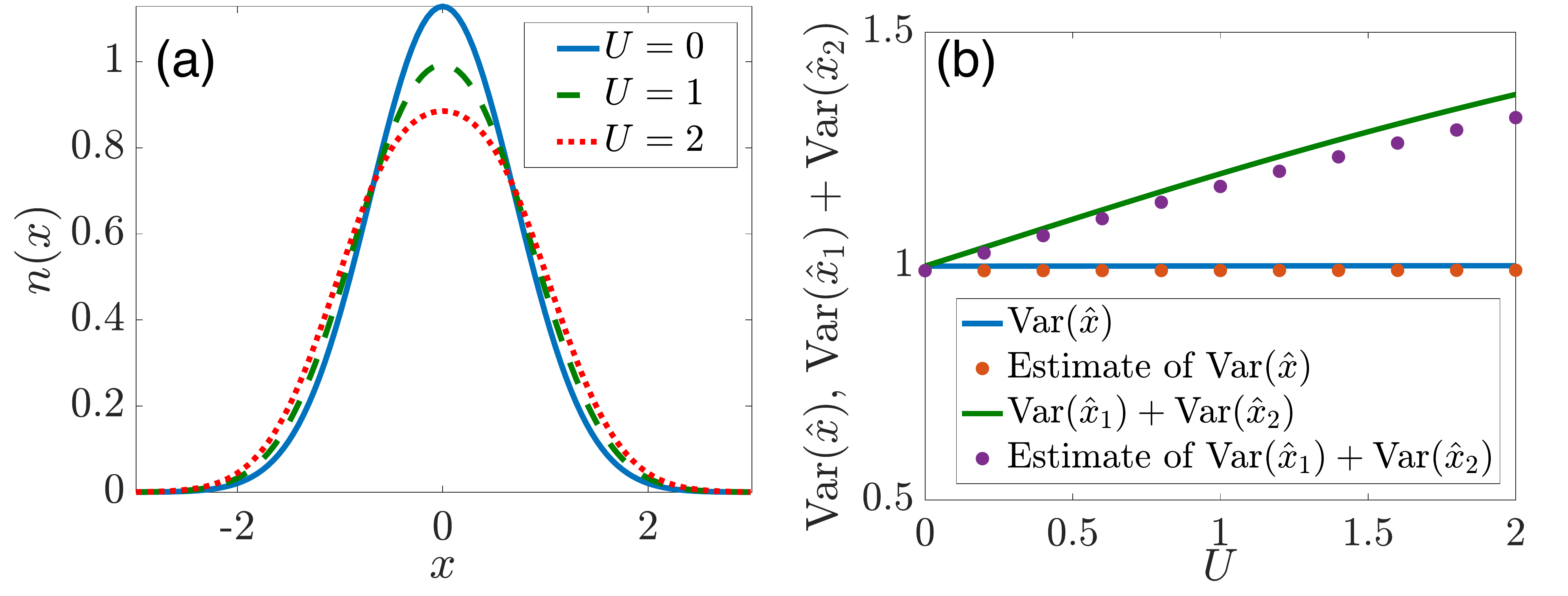}
\caption{Two interacting particles in a harmonic trap. (a) Density distribution $n(x)$ in the ground state, for increasing values of the interaction strength $U$ (in units of $\hbar \Omega\sqrt{\hbar/m\Omega}$); the position $x$ is expressed in units of $\sqrt{\hbar/m\Omega}$. (b) Spatial variances as extracted from integrated excitation rates (dots), and compared to their exact values (solid lines). The center-of-mass variance $\mathrm{Var}(\hat{x})$ is independent of the interaction strength $U$ (horizontal line). In contrast, the quantity $\mathrm{Var}(\hat{x}_1)\!+\!\mathrm{Var}(\hat{x}_2)$ captures well the spreading of the wave function upon increasing the repulsive interaction $U$ (tilted line). Simulations were performed with a modulation strength $E\!=\!0.01 \sqrt{m\hbar \Omega^3}$ and an observation time $t\!=\!5/\Omega$; the excitation rates were integrated over $\omega$ up to the value $10\Omega$, using  discrete steps of $0.05\Omega$.}
\label{interacting}
\end{center}
\end{figure}

\textit{Interacting particles in a harmonic trap.}
We now consider a system of two particles of mass $m$, moving in a one-dimensional harmonic trap of frequency $\Omega$ (see Ref.~\cite{Sup_Mat} for the single-particle case). We assume that the two particles are distinguishable~\cite{indistinguishable}, and that they interact via a repulsive contact interaction $U \delta (\hat{x}_1 - \hat{x}_2)$, with $U\!>\!0$. The interaction spreads out the ground-state wavefunction, as can be seen in the density distributions $n(x)$ depicted in Fig.~\ref{interacting}(a) (see Ref.~\cite{Busch:1998} for exact solutions). This spreading is experimentally relevant in ultracold-atom experiments realizing bosonic Mott insulators, where it was shown to affect the effective on-site interaction~\cite{Campbell:2006,Will:2010,Will:2011,Zurn:2012}. We now describe  how this delocalization through interactions could be finely resolved using excitation-rate measurements.

First, we note that the two-body Schr\"odinger equation can be decomposed in terms of the center of mass and relative motions. The center of mass is known to be independent of the interparticle interactions~\cite{Busch:1998}, and hence the related variance $\mathrm{Var}(\hat{x})\!=\!\mathrm{Var}(\hat{x}_1\!+\!\hat{x}_2)$ does not depend on $U$. In contrast, the density spread in Fig.~\ref{interacting}(a) is accurately captured by the relevant quantity $\mathrm{Var}(\hat{x}_1)\!+\!\mathrm{Var}(\hat{x}_2)\!=\!(1/2)\left [\mathrm{Var}(\hat{x}_1-\hat{x}_2)+\mathrm{Var}(\hat{x})\right ]$, which is associated with the relative motion and  depends on $U$. While $\mathrm{Var}(\hat{x})$ is directly accessible through the driving protocol described above [Eqs.~\eqref{driving_ham}--\eqref{integrated_rate}], the detection of $\mathrm{Var}(\hat{x}_1-\hat{x}_2)$ requires a particle-dependent modulation of the form $\hat V (t)\!=\!2E(\hat{x}_1 - \hat{x}_2) \cos (\omega t)$. Such a drive can be realized by considering two atomic internal states with opposite magnetic moments subjected to an oscillating magnetic field~\cite{Aidelsburger:2013,Kennedy:2013}. We describe below how this modification of the driving scheme allows for an accurate evaluation of the two-particle wave-function spreading.

We have numerically calculated the integrated excitation rates $\Gamma_{\hat{x}_1 - \hat{x}_2}^{\mathrm{int}}$ ($\Gamma_{\hat{x}_1 + \hat{x}_2}^{\mathrm{int}}$), upon subjecting the two-particle system to the particle-dependent (independent) modulations. From these excitation rates, we can estimate both $\mathrm{Var}(\hat{x})$ and $\mathrm{Var}(\hat{x}_1) + \mathrm{Var}(\hat{x}_2)$. The numerical results presented in Fig.~\ref{interacting}(b) show that the variance estimated from excitation-rate measurements (dots) perfectly reproduces the exact result (solid line).
These simulations confirm that the delocalization-by-interaction effect can be quantitatively measured in ultracold atoms through spectroscopic responses.

\textit{Localized spin excitations.} The localization of spin excitations, in (disordered) Heisenberg spin chains~\cite{Lyo,Igarashi,Pan}, can be detected using this same spectroscopic approach. The modulation $\hat V(t)\!=\!2E\cos(\omega t) \sum_j j \hat S_j^+ \hat S_j^-$ should be proportional to the center of mass of the magnonic excitation~\cite{Matsubara}, where $\hat S^{\pm}_j$ are spin operators on site $j$.

\textit{Many-body quantum geometric tensor.}
Dissipative responses are closely related to the concept of quantum geometry~\cite{Souza:2000,Kolodrubetz:PhysRep,Tran:2017,Tran:2018,OzawaGoldman, Asteria:2018}. 
At a fundamental level, the geometry of a quantum state $|\psi (\boldsymbol\theta) \rangle$, which depends on a set of parameters $\boldsymbol\theta\!=\!(\theta_1, \theta_2, \cdots)$, is described by the quantum geometric tensor~\cite{Kolodrubetz:PhysRep}
\begin{align}
	\chi_{ij}(\boldsymbol\theta) \equiv 
	\langle \partial_{\theta_i} \psi | \partial_{\theta_j} \psi \rangle
	-
	\langle \partial_{\theta_i} \psi |\psi \rangle
	\langle \psi | \partial_{\theta_j} \psi \rangle.\label{QGT_general}
\end{align}
Its imaginary part is related to the Berry curvature, $\mathrm{Im}\chi_{ij}\!=\!\Omega_{ij}/2$, which is associated with the physics of the geometric (Berry) phase and topological matter~\cite{Mead:RMP,Xiao:RMP, Hasan:RMP, Qi:RMP, Cooper:RMP, Ozawa:RMP}, whereas its real part is known as the quantum metric (or Fubini-Study metric) tensor~\cite{Provost:1980,Anandan:1990,Kolodrubetz:PhysRep}, $\mathrm{Re}\chi_{ij}\!=\!g_{ij}$. The full quantum geometric tensor was recently extracted from Rabi-oscillation measurements in diamond nitrogen-vacancy centers~\cite{Yu:2018}, from polarization tomography in polaritons~\cite{Gianfrate:2018}, and through similar methods in superconducting qubits~\cite{Tan:2019}.

The ground state of a many-body Hamiltonian can exhibit nontrivial geometric and topological properties. This can be revealed by introducing twisted boundary conditions~\cite{Niu:1985,Souza:2000,Watanabe:2018,Kudo:2019} and by calculating the quantum geometric tensor~\eqref{QGT_general} in the parameter space spanned by the corresponding twist angles $\boldsymbol\theta\!=\!(\theta_x,\theta_y, \dots)$. As shown in Ref.~\cite{Souza:2000}, the real part of the so-defined quantum geometric tensor, i.e., the many-body quantum metric, describes the variance of the position operator~\cite{Sup_Mat}
\begin{align}
	g_{xx}^\mathrm{MB}
	=
	\mathrm{Var}(\hat{x}) / L_x^2 , \label{gxxvarx}
\end{align}
where $L_x$ denotes the system's length along $x$. 

Combining Eq.~\eqref{gxxvarx} with Eq.~\eqref{integrated_rate} indicates that the many-body quantum metric $g_{xx}^\mathrm{MB}$ is  directly accessible through excitation-rate measurements ($\Gamma_{\hat{x}}^{\mathrm{int}}$). Similarly, the integrated rate $\Gamma_{\hat{y}}^{\mathrm{int}}$ upon linear shaking along the $y$ direction is proportional to $g_{yy}^\mathrm{MB}$; if the modulation is aligned along the diagonal $x + y$ direction, the resulting integrated rate is proportional to $L_x^2 g_{xx}^\mathrm{MB} + 2 L_x L_y g_{xy}^\mathrm{MB} + L_y^2 g_{yy}^\mathrm{MB}$.
Hence, all the components of the many-body quantum metric can be extracted from excitation-rate measurements upon linear shaking.

On the other hand, the many-body Berry curvature is related to the integrated rate upon circular shaking~\cite{Repellin:2018}. Indeed, considering the periodic modulation $\hat V_{\pm} (t)\!=\!2E [\hat{x} \cos (\omega t) \pm \hat{y} \sin (\omega t)]$, the integrated rates read
\begin{align}
	\Gamma_{\pm}^{\mathrm{int}}
	&=
	\frac{2\pi E^2}{\hbar^2}\sum_{\beta \neq \alpha}
	\left| \left\langle \beta \left|
	\hat{x} \mp i\hat {y}
	\right|\alpha \right\rangle \right|^2
	\notag \\
	&=
	\frac{2\pi E^2}{\hbar^2}
	\left( L_x^2 g_{xx}^\mathrm{MB} + L_y^2 g_{yy}^\mathrm{MB} \pm L_x L_y \Omega_{xy}^\mathrm{MB} \right).
\end{align}
Therefore, the many-body Berry curvature is given by the differential integrated rate per unit area: 
\begin{equation}
(\Gamma_+^\mathrm{int} - \Gamma_-^\mathrm{int})/2L_x L_y = \frac{2\pi E^2}{\hbar^2} \Omega_{xy}^\mathrm{MB}.
\end{equation}
This relation between circular dichroism and the many-body Berry curvature (or nonintegrated Chern number~\cite{Kudo:2019}) can also be derived from Kramers-Kronig relations~\cite{Repellin:2018}.

Summarizing, all the components of the many-body quantum geometric tensor are related to an observable response of the system upon linear or circular shaking. This result generalizes previous connections between the quantum geometry of single-particle states and spectroscopic responses~\cite{Tran:2017,Tran:2018, Asteria:2018, OzawaGoldman} to a many-body framework. In the special case of a band insulator, the many-body quantum metric is proportional to the average of the single-particle quantum metric over the entire Brillouin zone~\cite{Sup_Mat}.

\textit{Conclusion.}
This work proposed spectroscopic responses as a novel method to study localization in quantum many-body systems, offering a practical alternative to {\textit in situ} imaging. It is intriguing to observe that such excitation-rate measurements can extract information on both geometry and localization, two important concepts in condensed matter, through the extraction of the many-body quantum geometric tensor. An exciting perspective concerns the application of the present approach to explore the localization properties of many-body quantum states of interest, such as excitations of fractional quantum Hall liquids~\cite{Yoshioka:Book, Martin:2004}, many-body localized systems~\cite{Nandkishore:2015, Abanin:2018}, and fractons~\cite{Nandkishore:2019}.

\begin{acknowledgments}
We thank insightful discussions with C\'ecile Repellin, Monika Aidelsburger, and Yoshiro Takahashi.
T.O. was supported by JSPS KAKENHI Grant No. JP18H05857, RIKEN Incentive Research Project, and the Interdisciplinary Theoretical and Mathematical Sciences Program (iTHEMS) at RIKEN.
N.G. was supported by the FRS-FNRS (Belgium) and the ERC Starting Grant TopoCold.
\end{acknowledgments}

\clearpage

\begin{widetext}

\begin{center}
\begin{large}
{\bf Supplemental material for: \\ ``Probing localization and quantum geometry by spectroscopy"}
\end{large}
\end{center}

\end{widetext}

\section{Application:~A single particle in a harmonic trap}
To illustrate the principle of our method, we consider the simplest case of one particle in a one-dimensional harmonic trapping potential. The Hamiltonian is of the form
\begin{align}
	\hat{H}_0 = \frac{p^2}{2m} + \frac{1}{2}m\Omega^2 x^2.\label{1D_harmonic}
\end{align}
The eigenstates are labeled by an integer $n \ge 0$, and the eigenvalues are 
$\hat{H}_0 |n\rangle = \epsilon_n |n\rangle$, where $\epsilon_n = \hbar \Omega (n + 1/2)$.
We note that this example is already qualitatively different from earlier studies connecting quantum geometry to spectroscopic responses~\cite{Tran:2017,Tran:2018, OzawaGoldman, Asteria:2018}, in the sense that the Hamiltonian in Eq.~\eqref{1D_harmonic} lacks translational invariance. 

\begin{figure}[htbp]
\begin{center}
\includegraphics[width=5.0cm]{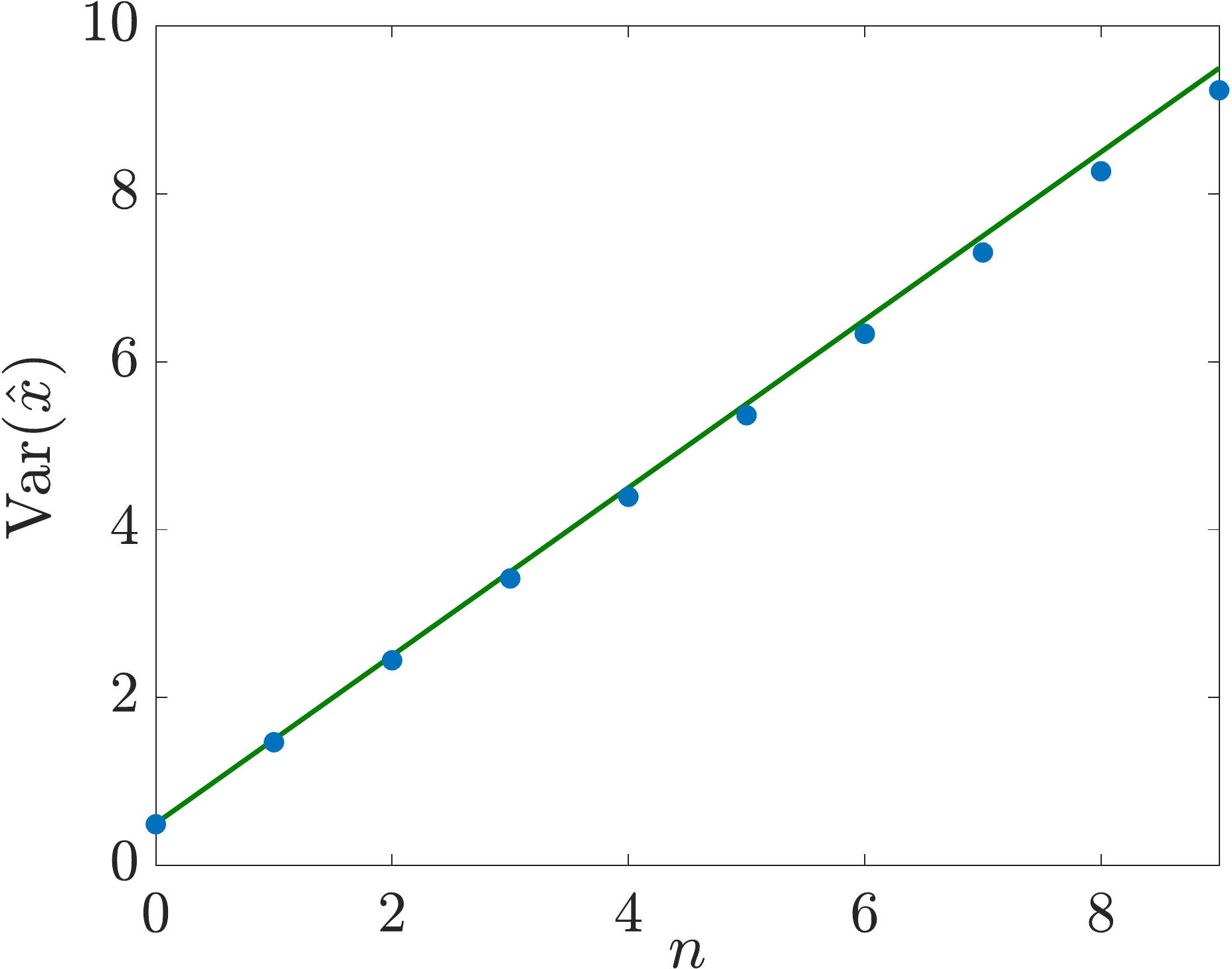}
\caption{Estimated $\mathrm{Var}(\hat{x})$ (dots) as a function of the initial state quantum number $n$. The length is measured in units of the oscillator length $\sqrt{\hbar/m\Omega}$. The line is the idea value $n+1/2$. We used $E = 0.01 \sqrt{m\hbar \Omega^3}$ and an observation time $t\!=\!5/\Omega$; the excitation rates were integrated over $\omega$ up to the value $4\Omega$, using  discrete steps of $0.05\Omega$.}
\label{simpleharmonic}
\end{center}
\end{figure}

We consider initializing the system in an arbitrary eigenstate $|n\rangle$. 
The variance of the position operator can be calculated analytically: $\mathrm{Var}(\hat{x}) = (n + 1/2)\hbar / m\Omega$.
We perform numerical simulations to calculate $\Gamma^\mathrm{int}_{\hat x}$, from which we estimate $\mathrm{Var}(\hat{x})$ and compare with the analytical result.
In the simulation, we start from a state $|n\rangle$ and we simulate the full time evolution under the drive, $\hat{H}_0 + 2E \cos (\omega t) \hat{x}$; we numerically estimate $\Gamma^{\mathrm{int}}$ by observing the probability of being in states other than $|n\rangle$ after some observation time.
In Fig.~\ref{simpleharmonic}, we plot the variance $\mathrm{Var}(\hat{x})$ estimated from the calculation of $\Gamma^\mathrm{int}$, for different values of $n$; the line corresponds to the analytical prediction, $(n + 1/2)\hbar / m\Omega$. The perfect agreement between our numerical results and the analytical prediction confirms the validity of our method.

\section{Derivation of Eq.(8)}
We hereby provide a derivation of Eq.(8) in the main text.
We start by considering the general $N$-body Hamiltonian
\begin{align}
	\hat{H}
	=
	\sum_{a=1}^N
	\left(
	\frac{[\hat{\mathbf{p}}_a - \mathbf{A}(\hat{\mathbf{r}}_a)]^2}{2m} + V(\hat{\mathbf{r}}_a)
	\right)
	+
	\hat{H}_\mathrm{int} (\{\mathbf{r}_a\}),\label{many_body_ham}
\end{align}
where $a$ labels the particles (bosons or fermions); the operators $\hat{\mathbf{p}}_a$ and $\hat{\mathbf{r}}_a$ are the momentum and position operators for the $a$-th particle; 
$\mathbf{A}(\hat{\mathbf{r}}_a)$ denotes a vector potential that is potentially coupled to the particles; $V(\hat{\mathbf{r}}_a)$ is a potential energy term (e.g.~a trapping or disordered potential). The inter-particle interaction is given by $\hat{H}_\mathrm{int}$, which we assume does not depend on momenta $\{\mathbf{p}_a\}$.

Following Niu-Thouless-Wu~\cite{Niu:1985} and Souza-Wilkens-Martin~\cite{Souza:2000}, we define the quantum geometric tensor in the space of twist angles of the boundary conditions. First, we note that twisted boundary conditions are mathematically equivalent to inserting magnetic fluxes through the system, while assuming  periodic boundary conditions. In this alternative description, the many-body Hamiltonian contains the inserted fluxes through the expression~\cite{Niu:1985}
\begin{align}
	\hat{H}(\boldsymbol\theta)
	=&
	\sum_{a=1}^N
	\left(
	\frac{[\hat{\mathbf{p}}_a - \mathbf{A}(\hat{\mathbf{r}}_a) + \hbar \boldsymbol\theta / \mathbf{L}]^2}{2m} + V(\hat{\mathbf{r}}_a)
	\right)
	+
	\hat{H}_\mathrm{int},\notag
\end{align}
where $\boldsymbol\theta\!\equiv\!(\theta_x, \theta_y, \cdots)$, and we introduced the notation $\boldsymbol\theta/\mathbf{L} \!\equiv\!(\theta_x/L_x, \theta_y/L_y, \cdots)$, where $L_i$ is the length along the $i$th direction. Gauging away this additional term in the vector potential reveals that $\theta_i$ indeed corresponds to the twist angle along the $i$th direction.

We consider an eigenstate $|\alpha\rangle$ of the many-body Hamiltonian $\hat{H} (\boldsymbol\theta)$; as in the main text, we assume that this state is non-degenerate and that it is well separated from all other states by spectral gaps. In general, the eigenstate depends on $\boldsymbol\theta$, and so does its energy, $\hat{H}(\boldsymbol\theta) | \alpha\rangle = \epsilon_\alpha (\boldsymbol\theta) | \alpha\rangle$.

We first take a commutator between $\hat{x}$ and $\hat{H}(\boldsymbol\theta)$:
\begin{align}
	\left[\hat{x}, \hat{H}(\boldsymbol\theta) \right]
	&=
	\sum_{a=1}^N
	\left[
	\hat{x}_a, \frac{[\hat{p}_{a,x} - A_x (\hat{\mathbf{r}}_a) + \hbar \theta_x/L_x]^2}{2m}
	\right]
	\notag \\
	&=
	i\hbar\frac{1}{m}\sum_{a=1}^N
	\left( \hat{p}_{a,x} - A_x (\hat{\mathbf{r}}_a) + \hbar \theta_x/L_x \right)
	\notag \\
	&=
	iL_x \partial_{\theta_x} \hat{H}(\boldsymbol\theta).
\end{align}
Taking the matrix element on both sides with different many-body eigenstates $|\alpha\rangle$ and $|\beta\rangle$, we obtain
\begin{align}
	(\epsilon_\beta (\boldsymbol\theta) - \epsilon_\alpha (\boldsymbol\theta))\langle \alpha | \hat{x} | \beta\rangle
	=
	iL_x \langle \alpha | \partial_{\theta_x} \hat{H}(\boldsymbol\theta) | \beta \rangle,
	\label{relation1}
\end{align}
where $\hat{H}(\boldsymbol\theta) | \alpha\rangle = \epsilon_\alpha (\boldsymbol\theta) | \alpha\rangle$ and $\hat{H}(\boldsymbol\theta) | \beta\rangle = \epsilon_\beta (\boldsymbol\theta) | \beta\rangle$.

On the other hand, taking a derivative of the equation $\langle \alpha | \hat{H}(\boldsymbol\theta) = \langle \alpha | \epsilon_\alpha (\boldsymbol\theta)$ with respect to $\theta_x$, and applying $|\beta \rangle$ from right, we obtain the following relation:
\begin{align}
	(\epsilon_\beta (\boldsymbol\theta) - \epsilon_\alpha (\boldsymbol\theta))
	\langle \partial_{\theta_x} \alpha|\beta\rangle = 
	-\langle \alpha | \partial_{\theta_x} \hat{H}(\boldsymbol\theta) | \beta \rangle.
	\label{relation2}
\end{align}
Comparing (\ref{relation1}) and (\ref{relation2}), we obtain
\begin{align}
	\langle \alpha | \hat{x} | \beta\rangle
	=
	-iL_x \langle \partial_{\theta_x} \alpha|\beta\rangle. \label{importantrelation}
\end{align}
Here we used that $|\alpha\rangle$ is spectrally separated so that $\epsilon_\beta (\boldsymbol\theta) \neq \epsilon_\alpha (\boldsymbol\theta)$. As previously noted in Ref.~\cite{Souza:2000}, while the position operator $\hat{x}$ is ill-defined upon applying periodic boundary conditions, Eq.~\eqref{importantrelation}  accurately describes its matrix elements. In this sense, this useful relation allows one to calculate physically meaningful quantities, such as localization, in closed geometries.

Using the relation (\ref{importantrelation}), we see
\begin{align}
	g_{xx}^\mathrm{MB}(\boldsymbol\theta)
	&=
	\langle \partial_{\theta_x} \alpha |\partial_{\theta_x} \alpha \rangle
	-
	\langle \partial_{\theta_x} \alpha |\alpha \rangle
	\langle \alpha |\partial_{\theta_x} \alpha \rangle
	\notag \\
	&=
	\sum_{\beta \neq \alpha} \langle \partial_{\theta_x} \alpha |\beta \rangle  \langle \beta |\partial_{\theta_x} \alpha \rangle
	\notag \\
	&=
	\sum_{\beta \neq \alpha} \langle \alpha |\hat{x} | \beta \rangle  \langle \beta |\hat{x}| \alpha \rangle
	/L_x^2
	=
	\mathrm{Var}(\hat{x})/L_x^2,
\end{align}
which is Eq.(8) in the main text.

\section{Relation between the many-body quantum metric and the quantum metric of Bloch bands}
We show how the many-body quantum metric $g^\mathrm{MB}_{ij} (\boldsymbol\theta)$ relates to the quantum metric of a non-interacting band structure $g^0_{ij}(\mathbf{k})$, when considering a Bloch band filled with non-interacting fermions. In this example, we take the occupied band to be the lowest one in the spectrum. According to the results presented in the main text, the following relation holds:
\begin{align}
	\Gamma_{\hat{x}}^\mathrm{int} = \frac{2\pi E^2}{\hbar^2} \sum_{\beta \neq \alpha}
	\left| \left\langle \beta \left|
	\hat{x}
	\right|\alpha \right\rangle \right|^2
	=
	\frac{2\pi E^2}{\hbar^2}
	L_x^2 g_{xx}^\mathrm{MB}(\boldsymbol\theta), \label{resultmaintext}
\end{align}
where $|\alpha\rangle$ corresponds to the completely filled band of fermions, and $|\beta\rangle$ is any other state.
We now examine the quantity $\sum_{\beta \neq \alpha}\left| \left\langle \beta \left| \hat{x} \right|\alpha \right\rangle \right|^2$, and rewrite it in terms of the quantum metric $g^0_{ij}(\mathbf{k})$ defined in the Brillouin zone of the non-interacting band structure.
Since $\hat{x}$ is a single-particle operator, if a state $|\beta\rangle$ contains more than one states outside the lowest band, the matrix element $\left\langle \beta \left| \hat{x} \right|\alpha \right\rangle$ vanishes. Therefore, for $|\beta\rangle$, we need to consider those states which are obtained by annihilating one state from $|\alpha\rangle$ and creating one state outside the lowest band. Let us write that $|\beta\rangle$ is obtained by annihilating a single-particle Bloch state $e^{i\mathbf{k}\cdot \mathbf{r}}|u_1 (\mathbf{k})\rangle/\sqrt{V}$ from $|\alpha\rangle$ and creating a Bloch state $e^{i\mathbf{k}^\prime\cdot \mathbf{r}}|u_n (\mathbf{k}^\prime)\rangle/\sqrt{V}$, where $n \neq 1$ is the band index for an excited band and $V$ is the total number of lattice sites. For such $|\beta\rangle$, we have~\cite{OzawaGoldman}
\begin{align}
	\left\langle \beta \left| \hat{x} \right|\alpha \right\rangle
	&=
	\langle u_n (\mathbf{k}^\prime) | e^{-i\mathbf{k}^\prime \cdot \mathbf{r}} \hat{x}_1 e^{i\mathbf{k}\cdot \mathbf{r}} |u_1(\mathbf{k}) \rangle
	\notag \\
	&=
	i\delta_{\mathbf{k},\mathbf{k}^\prime} \langle u_n(\mathbf{k})|\partial_{k_x} u_1 (\mathbf{k})\rangle.
\end{align}
When summing over $|\beta\rangle$, we need to sum over $n\neq 1$, $\mathbf{k}$, and $\mathbf{k}^\prime$. Then,
\begin{align}
	\sum_{\beta \neq \alpha}
	\left| \left\langle \beta \left|
	\hat{x}
	\right|\alpha \right\rangle \right|^2
	&=
	\sum_{\mathbf{k}} \sum_{n\neq 1}
	\left| \langle u_n(\mathbf{k})|\partial_{k_x} u_1 (\mathbf{k})\rangle \right|^2
	\notag \\
	&=
	\sum_{\mathbf{k}}g_{xx}^0 (\mathbf{k}).
\end{align}
Comparing this result with Eq.~\eqref{resultmaintext}, we eventually obtain the following relation
\begin{align}
	L_x^2 g_{xx}^\mathrm{MB}(\boldsymbol\theta) = \sum_{\mathbf{k}}g_{xx}^0 (\mathbf{k}).
\end{align}

In this non-interacting filled-band configuration, the integrated excitation rate upon linear driving 
reads
$\Gamma_{\hat{x}}^\mathrm{int} = (2\pi E^2/\hbar^2) \sum_{\mathbf{k}}g_{xx}^0 (\mathbf{k})$, which is indeed consistent with the result previously obtained in Ref.~\cite{OzawaGoldman}.


\begin{thebibliography}{99}
\bibitem{remark0} The results presented in this work are still valid for degenerate states, as long as the position operator does not couple different states within the degenerate manifold.
\bibitem{indistinguishable} In the case of indistinguishable particles, the protocol would give access to $\mathrm{Var}(\hat{x})$ through $\Gamma_{\hat{x}_1 + \hat{x}_2}^{\mathrm{int}}$.
\end{thebibliography}
\end{document}